**AttackPathGNN: Cross-function vulnerability detection in smart contracts using state interference graphs and conjunction pooling**

Gabriela DOBRIȚA, Simona-Vasilica OPREA*, Adela BÂRA
Bucharest University of Economic Studies, Department of Economic Informatics and Cybernetics, no.6 Piața Română, Bucharest, 010374, Romania, *Corresponding author: simona.oprea@csie.ase.ro

**Abstract:** Existing learning-based detectors for Solidity smart-contracts reduce vulnerability detection to syntactic pattern matching within single functions, yet many of the most consequential exploits (The DAO, Cream Finance) exist not in any individual function but in the relationship between functions and in the combination of conditions that made the attack feasible. Thus, we propose AttackPathGNN, a graph neural network (GNN) that reframes detection as reasoning over explicit attack paths. Two architectural choices distinguish it from prior GNN-based detectors: (1)a State Interference Graph that links every pair of functions sharing mutable storage through typed, weighted edges and through directed reentrancy-path edges defined by an explicit five-condition predicate; (2)conjunction pooling, a differentiable AND-aggregator over eight named exploit preconditions whose log-sigmoid form causes the per-function exploit score to collapse whenever any single mitigation (a reentrancy guard, an access-control modifier or SafeMath) is in place. Across five independent training runs, AttackPathGNN attains 92.3±0.2% F1 on the SmartBugs Wild held-out test partition (4.3±0.3% false-negative rate, 90.8±2.5% detection rate on the independently human-labelled SmartBugs Curated benchmark), recovering 6/10 DASP10 categories at 100% on every seed and Reentrancy at 98.7±1.8%. Each prediction is emitted with a structured remediation report, turning each verdict into an actionable, function-level audit finding.

## 1. Introduction

Smart contracts on Ethereum hold real value and a single line of vulnerable Solidity code can drain millions of dollars in seconds. Automated vulnerability detection has therefore become one of the most active research areas in blockchain security, with both static analysers and learning-based models proposed in rapid succession over the past five years. Yet the field has accumulated detectors faster than it has accumulated successful detections of the exploits that actually matter. The reason is structural. Most existing tools: symbolic engines such as Mythril [1] and Securify [2], regex-based linters such as SmartCheck [3] and the dominant family of GNN-based classifiers, treat vulnerability detection as a pattern-matching problem within a single function. They ask whether withdraw() contains a suspicious pattern or whether transfer() follows safe coding practices. If a function looks clean in isolation, it is declared safe. The trouble is that some of the most consequential exploits, including The DAO [4] and Cream Finance [5], never lived inside a single function, but in the relationship between functions and in the precise combination of conditions that made the attack feasible. Cross-function reentrancy, stale-balance accounting and shared-state races are invisible to a detector whose unit of analysis is one function at a time.

We therefore reformulate detection around two coupled research questions:

RQ1. *Which functions interact through shared state in a way that exposes an attack surface?*

RQ2. *Which conditions for exploitation hold simultaneously?*

A practical detector must answer both, and its output must tell a developer not just whether a contract is vulnerable, but exactly where the attack path runs and which single change would close it. In this context, we propose AttackPathGNN, a GNN that embodies this reformulation through two architectural choices. A State Interference Graph makes cross-function state dependencies first-class structure: every pair of functions sharing mutable storage is connected by a typed, weighted edge (write–read, read–write, write–write) and a directed reentrancy-path edge fires only when an explicit five-condition predicate holds. Conjunction pooling aggregates eight named exploit preconditions per function through a sum of log-sigmoid scores, a differentiable AND-aggregator whose log-product form collapses the per-function exploit score whenever any single mitigation is in place. Together, the two pieces give the model a structural account of where attacks can run and a probabilistic account of when they can succeed.

We treat classification accuracy as a baseline requirement rather than the central contribution. The central goal is a detector whose internal reasoning aligns with how exploits actually work, through chains of conditions over interacting functions, and whose output is structured enough to drive concrete remediation rather than another opaque binary label. We demonstrate this on three fronts. Across five independently seeded training runs, AttackPathGNN reaches a mean detection rate of 90.8±2.5 % on the human-labeled SmartBugs Curated benchmark [6] (best seed 94.4%, worst seed 88.1 %), with six DASP10 categories: arithmetic, denial of service, front-running, time manipulation, short address and "other", solved at 100% across every seed, reentrancy recovered at 98.7±1.8% and bad randomness at 95.0±6.9 %. The rule-only ablation of the same architecture, in which the eight precondition scores are thresholded directly without any GNN-learned components, recovers only 16.8 % of the same contracts, isolating a +74-percentage-point margin attributable entirely to the learned components.

This work makes four contributions:

(i) A heterogeneous graph representation combining weighted, typed state-interference edges decomposed by overlap kind (write–read, read–write, write–write), directed reentrancy-path edges defined by an explicit five-condition predicate and first-class precondition nodes attached to each function.

(ii) Conjunction pooling, a differentiable operator that aggregates per-function precondition scores through a sum of log-sigmoid terms, enforcing AND-logic over the exploit chain end-to-end. To our knowledge no prior smart-contract detector uses a differentiable conjunction over named exploit preconditions in this form.

(iii) A structured interpretability output consisting of ranked dangerous function pairs with shared variable names and interference type, per-function precondition satisfaction profiles and the specific precondition whose mitigation would close the attack path. The output is named, programmatically aggregable and directly maps to a concrete remediation suggestion, going beyond line-level localization [7] or attention heatmaps [8].

(iv) Empirical evidence of generalization, reported as the mean and standard deviation over five independently seeded training runs, a level of statistical reporting absent from prior GNN-based smart-contract detectors.

## 2. Literature review

### 2.1 Smart contract vulnerabilities and security challenges

The growth of decentralized finance (DeFi) has increased the importance of smart contract security due to the substantial financial losses associated with vulnerabilities. A comprehensive assessment of 127 real-world attacks showed that existing automated security tools could have prevented only 8% of the analyzed incidents, highlighting the limitations of current approaches in addressing logic-related and protocol-level vulnerabilities [9], demonstrating the need for more advanced and context-aware security solutions capable of detecting complex attack vectors beyond traditional vulnerability classes.

Several studies have focused on detecting specific vulnerabilities in deployed smart contracts. Sereum introduced a runtime monitoring framework capable of protecting already deployed Ethereum contracts against reentrancy attacks with a false positive rate of only 0.06% [10]. Clairvoyance proposed a static analysis approach based on path feasibility checking to improve the detection of reentrancy vulnerabilities while reducing false positives [11]. Similarly, SAILFISH combined lightweight exploration with symbolic evaluation to identify state-inconsistency vulnerabilities such as reentrancy and transaction-order dependence, outperforming several established analyzers [12]. MAIAN introduced inter-procedural symbolic analysis to detect trace vulnerabilities spanning multiple contract invocations, identifying thousands of exploitable contracts in large-scale Ethereum deployments [13]. Subsequent empirical analyses revealed that only a small fraction of reported vulnerable contracts are actually exploited in practice, highlighting the gap between theoretical and real-world risk [14]. Gas-related vulnerabilities have also received attention, with eTainter employing taint analysis to identify denial-of-service risks caused by excessive gas consumption, achieving superior precision and recall compared to previous methods [15].

### 2.2 Deep learning and GNNs for vulnerability detection

One line of research constructs contract graphs that capture both syntactic and semantic structures of smart contract code. Using a Degree-Free Graph Convolutional Network (DR-GCN) and a Temporal Message Propagation (TMP) network, researchers demonstrated significant improvements over state-of-the-art vulnerability detection methods [16]. DA-GNN enhanced vulnerability detection by introducing dual attention mechanisms that jointly model semantic and relational information within control-flow graphs [17]. HGAT further improved detection accuracy through hierarchical graph attention networks that integrate abstract syntax trees and control-flow information [18].

Further developments integrated expert knowledge with graph-based learning. By combining control-flow and data-flow information with handcrafted security patterns, a hybrid GNN framework achieved detection accuracies of 89.15%, 89.02%, and 83.21% for reentrancy, timestamp dependence and infinite loop vulnerabilities, respectively [7]. To further improve representation learning, Peculiar introduced a pre-training strategy based on crucial data-flow graphs, achieving precision and recall values exceeding 91% for reentrancy vulnerability detection across more than 40,000 smart contracts [8]. Recent graph pooling methods based on differentiable minCUT optimization have further improved hierarchical graph representation learning and clustering efficiency [19].

### 2.3 Heterogeneous graph learning for smart contract analysis

The use of heterogeneous graph representations has emerged as a promising direction for smart contract analysis. MANDO introduced a heterogeneous graph representation that combines control-flow and call graphs through customized meta-path extraction and attention mechanisms. The framework significantly improved vulnerability detection performance and became the first learning-based method capable of fine-grained line-level vulnerability identification [20]. Building upon this work, MANDO-GURU extended heterogeneous graph attention networks for both contract-level and line-level vulnerability detection. Experimental results demonstrated improvements of up to 24% in F1-score at the contract level and up to 63.4% at the line level compared with traditional code-analysis approaches [21]. Recent GNN-based frameworks have combined source-code and bytecode representations to achieve highly accurate detection across multiple vulnerability categories [22]. MCR-VD introduced an interpretable code-property-graph framework capable of localizing vulnerabilities at the line level while maintaining high detection performance [23].

### 2.4 Explainable AI for graph-based security models

As GNNs become increasingly complex, explainability has become a critical requirement for security applications. GNNExplainer was proposed as the first model-agnostic framework capable of identifying compact subgraphs and relevant node features responsible for GNN predictions, improving explanation accuracy by up to 43% over competing methods [24]. Similarly, SubgraphX addressed the explainability challenge by identifying important graph substructures using Monte Carlo Tree Search and Shapley values. By explicitly modeling the contribution of subgraphs to predictions, the method produced more intuitive and human-understandable explanations while maintaining computational efficiency [25]. GAVulExplainer extended explainable vulnerability detection by employing genetic algorithms to generate interpretable subgraph explanations for GNN predictions [26].

### 2.5 Neuro-symbolic learning and logical reasoning

Logic Tensor Networks (LTNs) combine neural networks with first-order fuzzy logic, enabling learning from noisy data while incorporating logical constraints. Experimental results demonstrated that logical background knowledge improves classification performance and robustness [27]. DeepProbLog further advanced neuro-symbolic learning by integrating deep neural networks into probabilistic logic programming. The framework supports symbolic reasoning, probabilistic inference, program induction, and end-to-end learning within a unified architecture [28]. Complementary research proposed differentiable neural-symbolic frameworks where neurons correspond to logical formulas, improving interpretability and resilience to inconsistent knowledge [29].

Another stream of research investigated differentiable fuzzy logic systems for weakly supervised learning. The findings revealed that many traditional fuzzy operators are unsuitable for gradient-based optimization and introduced new logical operators that better support learning in neural environments [30].

### 2.6 Transformer-based vulnerability prediction

Prior to transformer-based approaches, convolutional neural networks demonstrated promising results for vulnerability detection through automated feature extraction from source code [31]. Devign pioneered graph-based vulnerability detection by combining code semantic graphs with neural learning, significantly outperforming traditional machine-learning methods [32]. Transformer architectures have recently been adopted for software vulnerability prediction. LINEVUL introduced a Transformer-based approach capable of predicting vulnerabilities at both function and line levels. Evaluations on more than 188,000 real-world C/C++ functions showed substantial improvements over previous graph-based methods, including significantly higher F1-scores, improved line-level localization accuracy and reduced inspection effort [33].

### 2.7 Research gaps and contributions

Despite recent advances in smart contract vulnerability detection, several important limitations remain. Traditional static-analysis tools such as Securify, SmartCheck, Clairvoyance and eTainter primarily rely on predefined rules and vulnerability signatures, limiting their ability to identify complex attack scenarios involving interactions across multiple functions. Similarly, runtime monitoring approaches such as Sereum focus on detecting specific attack classes, particularly reentrancy, rather than providing a generalized framework for exploit reasoning.

Recent graph-based methods, including DR-GCN, DA-GNN, HGAT, MANDO, MANDO-GURU and MCR-VD, have significantly improved feature extraction by modeling syntactic, semantic, control-flow, and data-flow relationships within smart contracts [16–23]. However, these approaches predominantly treat vulnerability detection as a node- or graph-classification problem and largely focus on identifying suspicious code patterns. Consequently, they do not explicitly model the attack paths through which vulnerabilities emerge, nor do they capture the interaction between multiple exploit preconditions distributed across different functions.

Another limitation of existing learning-based detectors is that vulnerabilities are typically evaluated independently at the function level. Many real-world exploits, including The DAO and Cream Finance incidents, arise from state interference between functions, shared storage dependencies and the simultaneous satisfaction of multiple exploit conditions. Existing graph representations do not explicitly encode these attack-enabling relationships, which can lead to false positives when vulnerable patterns are present but appropriate mitigations exist.

To address these limitations, we propose AttackPathGNN, which reframes vulnerability detection as reasoning over explicit attack paths rather than isolated code patterns. First, we introduce a State Interference Graph that models dependencies between functions sharing mutable storage and captures reentrancy paths through explicitly defined attack relationships. Second, we propose conjunction pooling, a differentiable AND-aggregation mechanism that directly models exploit preconditions and allows vulnerability scores to collapse when any required condition is absent or mitigated. Unlike previous GNN-based detectors, AttackPathGNN explicitly represents both inter-function dependencies and exploit feasibility, thereby providing a closer approximation to how real-world attacks occur.

Furthermore, while recent explainable approaches such as GNNExplainer, SubgraphX and GAVulExplainer focus on post-hoc explanations of model predictions, AttackPathGNN produces structured remediation reports directly from the attack-path reasoning process, transforming vulnerability predictions into actionable audit findings. This combination of attack-path modeling, exploit-precondition reasoning, and actionable explanations represents a novel contribution to smart contract security analysis.

## 3. Methodology

Unlike conventional detectors that flag syntactic patterns independently per function, AttackPathGNN constructs a heterogeneous program graph in which cross-function state dependencies, reentrancy paths, and exploit preconditions are first-class structural elements. The system is organized as a seven-stage pipeline that transforms raw source code into a binary verdict (vulnerable/clean) accompanied by a structured explanation identifying the affected functions, the shared state variables and the precondition chain that enables the attack.

We use the following notation throughout. A contract $C$ contains a set of functions $\mathcal{F} = \{f_1, \dots, f_F\}$ and a set of persistent storage variables $\mathcal{S}$. For each function $f_i$, we write $R_i, W_i \subseteq \mathcal{S}$ for the storage variables it reads and writes, $X_i$ for the set of external calls it issues and $\delta_i \in \{0,1\}$ for an indicator that is 1 when at least one storage write occurs after the last external call (a checks–effects–interactions (CEI) violation). Three boolean flags $\mathrm{pub}(f_i), \mathrm{guard}(f_i), \mathrm{acl}(f_i) \in \{0,1\}$ record whether $f_i$ is publicly callable, carries a reentrancy-guard modifier and an access-control modifier respectively. We write $\sigma$ for the sigmoid function, $\|$ for vector concatenation and $\langle\cdot\rangle$ for an ordered tuple.

### 3.1 Pipeline overview

The methodology is organized as a sequential seven-stage pipeline, summarized in Table 1 and Figure 1. Each stage builds on the outputs of the previous one, and the pipeline as a whole transforms a contract $C$ into a binary classification together with a structured interpretability report.

Table 1. Pipeline stages of AttackPathGNN

| Stage | Component | Input | Output |
|---|---|---|---|
| **S1** | Source parsing (Slither AST / regex) | Raw Solidity source | Function signatures, storage read/write sets, modifiers, visibility |
| **S2** | Semantic pattern extraction | Source code / AST | Pattern node set; 30-dim feature vector padded to 35 |
| **S3** | State interference analysis | Per-function $R_i, W_i, X_i, \delta_i$ | Typed interference edges; directed reentrancy-path edges |
| **S4** | Precondition scoring | Function metadata from S1–S3 | Eight scores $s_{i,k} \in [0,1]$ per function |
| **S5** | Heterogeneous graph construction | Outputs of S1–S4 | Graph $G$ with four node types and seven edge relations |
| **S6** | GNN forward pass | Graph $G$ | Function-level exploit embeddings |
| **S7** | Classification and explanation | Pooled graph embedding | Binary verdict and per-function risk attribution |

For every stage, we state the motivation (why it is needed), the formal definition (what it computes), and the role its output plays downstream.

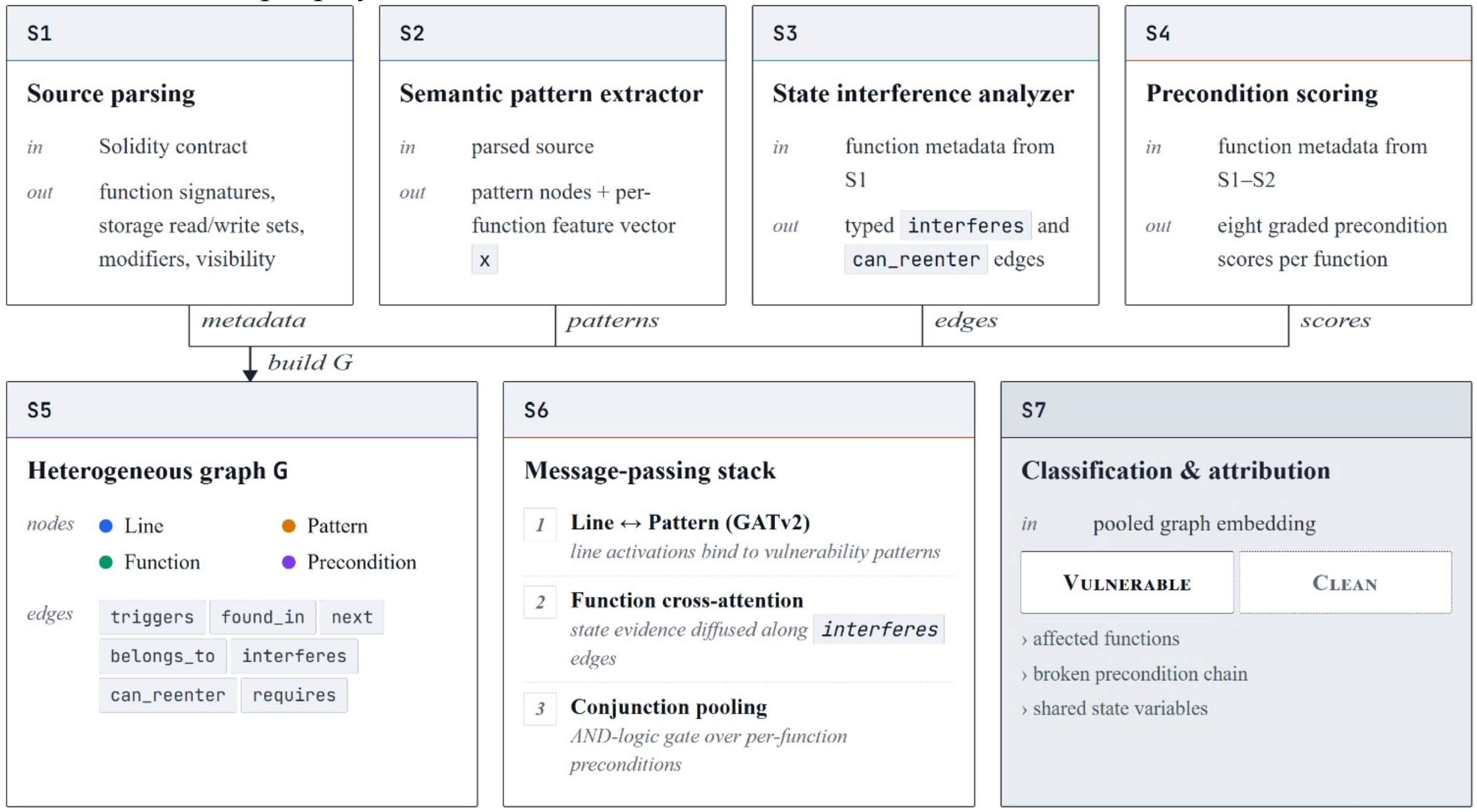

Figure 1. AttackPathGNN pipeline.

### 3.2 Stage 1-Source code parsing

The first stage converts raw Solidity source into a structured representation suitable for downstream graph construction. Two parsing strategies are employed and selected automatically based on tool availability.

*3.2.1 Primary path: Slither AST parsing*

When *slither-analyzer* is installed, the contract source is parsed into an Abstract Syntax Tree (AST) with resolved type information. Slither distinguishes persistent storage variables from ephemeral memory allocations, which is critical. A false positive arises precisely when a tool flags a local memory variable as a storage-write vulnerability. For every function, the AST yields its visibility (public, external, internal or private), its mutability (payable, view, pure), its modifier list (with explicit detection of reentrancy guards such as *nonReentrant* or *locked* and access-control modifiers such as *onlyOwner*; when no modifier-based access control is found, we additionally check for *require(msg.sender ...)* patterns inside the function body), its parameter types, its external calls classified by kind (*call*, *send*, *transfer*, *delegatecall*) and its source-line range. For every function with at least one external call, Slither also compares the source positions of the call nodes against those of state-write nodes. If any storage write occurs after the last external call, the flag $\delta_i$ is set to 1, marking a checks–effects–interactions violation. Constructors, fallback functions and *receive* functions are excluded from extraction because they are not part of the externally invocable surface.

*3.2.2 Fallback path: regex parsing*

When Slither is unavailable or fails to compile the contract (for example, due to missing imports or unsupported compiler versions), a regex-based parser is used as a fallback. Function signatures are matched against a single regular expression whose modifier capture group is bounded to 300 characters to prevent catastrophic backtracking on complex signatures. Function bodies are extracted by a brace-depth counter starting at the opening brace. Storage reads are extracted from mapping-indexed access expressions (e.g., *balances[addr]*), storage writes from compound and direct assignments ($x$ = ..., $x$ += ...), external calls from the standard *.call*, *.send*, *.transfer* and *delegatecall* patterns. Because Solidity allows local variables to shadow storage variables, the regex fallback may include local names in $R_i$ or $W_i$ as noise, addressed by the Slither-first design.

Both paths produce the same per-function schema: name, parameters, modifiers, body text, line range and boolean flags for visibility, payability, reentrancy guard and access control, so all downstream stages are parser-agnostic.

**3.3 Stage 2-Semantic pattern extraction**

The second stage extracts a contract-level feature vector of vulnerability-relevant pattern counts and structural metrics. Thirteen syntactic patterns are detected by compiled regular expressions: reentrancy via *.call{value: ...}*, *.send* and *.transfer*, state-after-call (CEI) violations, *tx.origin* use, unprotected *selfdestruct*, unchecked arithmetic operators (zeroed when SafeMath or Solidity ≥ 0.8 is detected), explicit overflow and underflow risk patterns (++, +=, --, -=), *delegatecall* invocations, unchecked low-level *call* returns, *block.timestamp* dependency in conditions and dependency on miner-controlled fields (*block.number*, *block.difficulty*, *block.coinbase*). These thirteen pattern counts are supplemented by seventeen structural counts (number of state variables, mappings, total functions, public and external functions, payable functions, modifiers, require and assert statements, lines of code, contracts, transfers inside loops, calls inside loops, and others), yielding a thirty-dimensional feature vector padded to thirty-five dimensions for the classifier input.

These features serve as baseline signals. The contribution of AttackPathGNN lies not in *which* patterns are detected but in *how* they are combined with the cross-function structure introduced in Stages 3–6. For each active pattern the contract also receives a *pattern node* in the heterogeneous graph (Stage 5), connected to the source line that triggered it.

**3.4 Stage 3-State interference analysis**

Prior GNN-based detectors construct graphs within individual functions or at the statement level. Cross-function state dependencies, through which real exploits propagate, are not explicitly modelled as graph structure. Stage 3 supplies precisely this missing structure.

*3.4.1 Storage access profile*

For every function $f_i \in \mathcal{F}$, we collect the four-tuple:

$$\mathcal{A}(f_i) = \langle R_i,\ W_i,\ X_i,\ \delta_i \rangle \tag{1}$$

with Slither, the sets $R_i$ and $W_i$ contain only persistent storage variables. With the regex fallback they may include local names, which increases the false-edge rate of the interference graph defined below.

*3.4.2 Interference relation*

Two distinct functions $f_i, f_j$ $(i \neq j)$ *interfere* if they share at least one mutable storage variable. We decompose the interference into three sub-types, capturing whether the overlap is between writes of $f_i$ and reads of $f_j$, reads of $f_i$ and writes of $f_j$ or writes on both sides:

$$I(f_i, f_j) = (W_i \cap R_j) \cup (R_i \cap W_j) \cup (W_i \cap W_j). \quad (2)$$

An *interference edge* $(f_i, f_j) \in E_{\text{interferes}}$ is created in the heterogeneous graph if and only if $I(f_i, f_j) \neq \emptyset$. Each such edge carries a four-dimensional feature vector consisting of the cardinalities of the three intersections in eq. (2) together with $|I(f_i, f_j)|$, projected to the model's hidden dimension $d$ by a learned linear map. The downstream cross-function attention layer (Stage 6) uses these features to weight interference edges by the *kind* and *extent* of overlap, distinguishing benign sharing (read-only or trivially small) from dangerous interaction.

*3.4.3 Reentrancy-path predicate*

A stricter, *directed* edge models the specific structural conditions required for a reentrancy attack path. We define the predicate:

$$\text{Reenter}(f_i \to f_j) \iff |X_i| > 0 \land \text{pub}(f_i) \land \neg\,\text{guard}(f_i) \land \text{pub}(f_j) \land \left(W_i \cap (R_j \cup W_j)\right) \neq \emptyset. \quad (3)$$

A directed edge $(f_i \to f_j) \in E_{\text{can_reenter}}$ is created exactly when eq. (3) holds. The attack semantics are explicit. An attacker calls $f_i$, which makes an external call, then control returns to attacker code, which re-enters via $f_j$ before $f_i$'s writes complete. Because $f_i$ writes a variable that $f_j$ reads or writes, the re-entry observes (or further mutates) inconsistent state. The predicate is intentionally asymmetric, capturing the *direction* of the attack, and intentionally local, adding a single guard modifier to $f_i$ flips $\text{guard}(f_i)$ and removes every outgoing reentrancy edge from $f_i$, providing a clean structural account of why a one-line fix neutralizes a vulnerability.

Reentrancy guards are detected by scanning each function's modifier list for names matching *nonReentrant*, *reentrancy* (case-insensitive substring) or *locked*. Access control is detected via modifier names (*onlyOwner*, *onlyAdmin*, role-based modifiers) or by *require(msg.sender ...)* checks in the function body. These detections feed both the edge construction above and the precondition extraction in Stage 4.

**3.5 Stage 4-Exploit precondition extraction**

A vulnerability *pattern* is not yet an *exploit*. An exploit requires every link in a precondition chain to hold simultaneously: an attacker-reachable entry point, a control-transfer instruction, an unguarded re-entry surface, a stale state read and so on. Breaking any single link, adding *nonReentrant*, restricting visibility, introducing access control, switching to checked arithmetic, renders the entire attack path infeasible. Stage 4 formalizes this chain as a set of differentiable precondition nodes attached to each function.

*3.5.1 The eight named preconditions*

For each function $f_i$, we evaluate eight preconditions $c_{i,k}$, $k = 1, \dots, 8$, each producing a satisfaction score $s_{i,k} \in [0,1]$ (Table 2). The binary-valued preconditions take values in {0,1}, while the two graded preconditions ($c_5$ and $c_8$) are clipped to $[0,1]$. A score of 1 means the precondition is *satisfied* (this link of the attack chain is intact), a score of 0 means the link is *broken*.

Table 2. The eight exploit preconditions

| ***k*** | **Precondition** | **Definition** |
|---|---|---|
| 1 | externally callable | $\text{pub}(f_i) = 1$ |
| 2 | has external call | $\|X_i\| \geq 1$ |
| 3 | state after call | $\delta_i = 1$ (CEI violation) |
| 4 | no reentrancy guard | $\text{guard}(f_i) = 0$ |
| 5 | shares mutable state | $\min\left(1,\ 1/5 \sum_{j \neq i,\, \text{pub}(f_j)} \|W_i \cap R_j\|\right)$ |
| 6 | no access control | $\text{acl}(f_i) = 0$ |
| 7 | handles value | $f_i$ is *payable*, uses *msg.value* or accesses *.balance* |
| 8 | unchecked arithmetic | $\min(1,\ 1/10 \cdot \#\{\text{unchecked arithmetic ops}\})$ |

*3.5.2 Precondition node embedding*

Each precondition $(i,k)$ becomes a node in the heterogeneous graph. Its initial embedding combines a learned type embedding for the precondition class with the scalar satisfaction score:

$$h_{c,(i,k)}^{(0)} = W_c\left[E_{\text{type}}[k] \,\|\, s_{i,k}\right], \quad (4)$$

where $E_{\text{type}} \in \mathbb{R}^{8\times d}$ is a learnable type-embedding matrix whose $k$-th row carries the *semantic identity* of the precondition class, $\|$ denotes concatenation and $W_c \in \mathbb{R}^{d\times(d+1)}$ projects the resulting $(d+1)$ dimensional vector back to the hidden dimension $d$. The scalar score is included as direct supervision. A high score signals to the GNN that this link of the exploit chain is intact and should propagate risk forward.

### 3.6 Stage 5-Heterogeneous graph construction

All extracted information is assembled into a single heterogeneous graph $G = (V, E, \tau, \phi)$, where $\tau$ assigns node types and $\phi$ assigns edge relation types.

*3.6.1 Node types*

The graph contains four node types (Table 3). All node embeddings have hidden dimension $d = 256$.

Table 3. Node types

| Type | Count | Initialization |
|---|---|---|
| Line | $L$ (one per source line) | Mean-pooled token embeddings (vocabulary 10 000, max 50 tokens per line) |
| Pattern | $P$ (one per active pattern) | Learned pattern-type embedding |
| Function | $F$ (one per function) | Linear projection of a 13-dim hand-crafted feature vector |
| Precondition | $8F$ (eight per function) | Type embedding plus scalar score, per (4) |

The function-node feature vector contains $[\text{pub}, \text{payable}, \text{view}, \text{guard}, \text{acl}, |X_i|, |R_i|, |W_i|, \delta_i, n_{\text{lines}}, n_{\text{cond}}, n_{\text{loop}}, \mathbb{1}[\text{selfdestruct}]]$.

*3.6.2 Edge relations*

Seven edge relations connect the node types (Table 4). The three relations, *interferes*, *can_reenter* and *requires*, are the structural backbone of AttackPathGNN's attack-path reasoning. Without them the model degenerates to a code-pattern detector at parity with prior GNN baselines.

Table 4. Edge relations

| Relation | Direction | Semantics |
|---|---|---|
| *triggers* | line → pattern | A source line activates a vulnerability pattern |
| *found_in* | pattern → line | Reverse lookup of *triggers* |
| *Next* | line → line | Sequential adjacency within a function body |
| *belongs_to* | line → function | A line is contained in a function |
| *interferes* | function ↔ function | Shared mutable storage, eq. (2) |
| *can_reenter* | function → function | Feasible reentrancy path, eq. (3) |
| *requires* | function→ precondition | Function $f_i$ owns precondition node $(i,k)$ |

*3.6.3 Complexity*

The heterogeneous graph contains $L + P + F + 8F = O(L + F)$ nodes and $O(LP + LF + F^2)$ edges. The pairwise function comparison driving *interferes* and *can_reenter* is $O(F^2|\mathcal{S}|)$, which is negligible in practice because $F < 50$ for virtually all Solidity contracts. The dominant cost of the pipeline is therefore the GNN forward pass itself, $O(|E|d^2)$, which scales linearly with the number of edges and quadratically with the hidden dimension.

### 3.7 Stage 6-GNN forward pass

The GNN consumes the heterogeneous graph and produces a fixed-size classification vector. The forward pass interleaves *standard* code-level message passing, adapted from established graph architectures, with a small number of *novel* operations specific to attack-path reasoning.

*3.7.1 Code-level message passing*

At the code level, the model first propagates information bidirectionally between line and pattern nodes through two GATv2 layers [34] over the *triggers* and *found_in* edges. A single GCN layer [35] over *next* edges then captures multi-line patterns spanning consecutive statements (for example, an external call on line $i$ followed by a state write on line $i+1$, a two-line CEI signature that no single-line pattern can express). A final GATv2 layer over *belongs_to* aggregates line representations into per-function embeddings, with the function node's hand-crafted feature vector serving as the attention query.

*3.7.2 Cross-function reasoning*

The function representations then pass through two further GATv2 layers, the first over the *typed, weighted interferes* edges (using the four-dimensional edge features in subsection 3.4.2), the second over the *directed can_reenter* edges. The first propagates the structural fact that two functions share mutable storage, while the second propagates *attacker reachability*. If $f_j$ is reachable from $f_i$ along a feasible reentrancy path, $f_j$'s representation is updated to reflect that an attacker entering through $f_i$ can reach it. We denote the resulting per-function representation $h_{f,i}^{(3)}$ and pass it to the conjunction-pooling step below. The attention weights produced by the *interferes* layer are retained and exposed to the interpretability stage as *dangerous interaction scores*.

*3.7.3 Conjunction pooling*

Standard pooling operators (sum, mean, max) implement OR-like aggregation. The output is high whenever *any* input is high. Exploit feasibility, however, requires every precondition to hold *simultaneously*, that is a single broken link in the chain (a reentrancy guard, an access-control modifier, SafeMath) must collapse the entire score, regardless of how many other risk factors remain active. We formalize this AND-logic with a *conjunction pooling* operator. Writing $g: \mathbb{R}^d \to \mathbb{R}$ for a learned two-layer Multilayer Perceptron (MLP) gate shared across precondition types, the per-function exploit score is:

$$\Phi(f_i) \;=\; \sum_{k=1}^{8} \log \sigma\left(g\left(h_{c,(i,k)} \,\|\, h_{f,i}^{(3)}\right)\right). \tag{5}$$

Figure 2 illustrates this collapse property empirically, showing the per-function precondition gate probabilities $\sigma\left(g\left(h_{c,(i,k)} \,\|\, h_{f,i}^{(3)}\right)\right)$ produced by AttackPathGNN on a synthetic vulnerable contract (left) and on a patched contract with nonReentrant modifiers added (right). In the left panel, every function carries an intact attack chain, the no_reent_guard column is uniformly red, and the per-function exploit scores are large. In the right panel the same column flips to green and by the log-sigmoid form of eq. (5), every per-function exploit score collapses to $-\infty$ regardless of the values of the remaining seven gates.

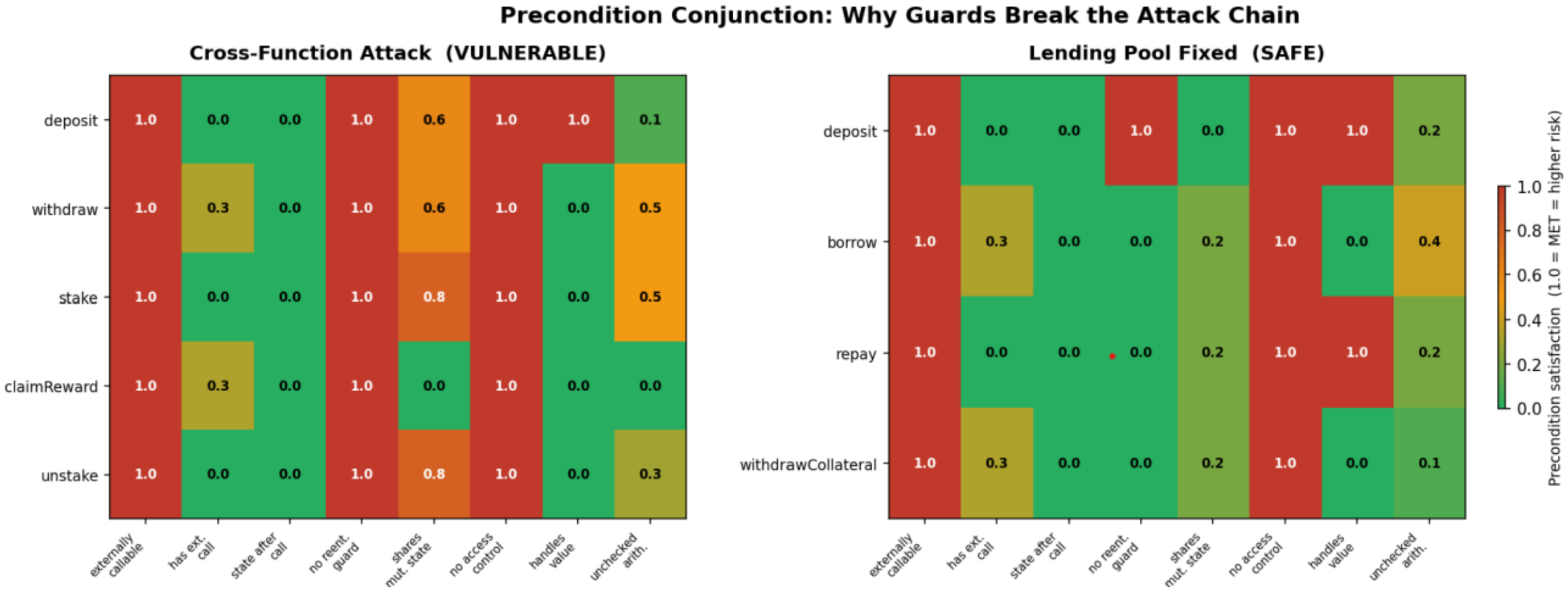

Figure 2. Per-function precondition gate probabilities on two synthetic Solidity contracts modelled after canonical exploit patterns.

The gate $g$ takes as input the concatenation ($\|$) of each precondition embedding $h_{c,(i,k)}$ with its parent function's interaction-aware representation $h_{f,i}^{(3)}$ (the output of the cross-function reasoning step), so that the model's belief in any one precondition can depend not only on what the precondition itself records, but also on the structural context of the function it belongs to. Eq. (5) admits a clean probabilistic interpretation. Writing $P(c_{i,k}) := \sigma\left(g(h_{c,(i,k)} \,\|\, h_{f,i}^{(3)})\right)$ for the model's belief that precondition $c_{i,k}$ is satisfied, and treating those eight beliefs as conditionally independent given the function's representation $h_{f,i}^{(3)}$, eq. (5) is exactly the log-probability that the entire precondition chain holds:

$$\Phi(f_i) = \log \prod_{k=1}^{8} P\left(c_{i,k}\right) = \log P\left(\bigwedge_{k=1}^{8} c_{i,k}\right). \tag{6}$$

The AND-logic property follows directly from the log-product form of eqs. (5) and (6): because $\log\sigma(z) \leq 0$ for all $z$ and $\log\sigma(z) \rightarrow -\infty$ as $\sigma(z) \rightarrow 0$, any single gate driven near zero, for example by a nonReentrant modifier setting $s_{i,4} = 0$ and pushing $g(h_{c,(i,4)} \parallel h_{f,i}^{(3)})$ toward $-\infty$, drives $\Phi(f_i) \rightarrow -\infty$, regardless of the values of the remaining seven gates. This is the differentiable analogue of the statement *"this function is not exploitable, because its reentrancy guard is active, even though every other condition for exploitation is met"*. The conjunction score is then fused with the function representation through a learned MLP, producing the final per-function embedding used downstream.

*3.7.4 Global pooling and classification*

A contract-level representation is formed by concatenating mean- and max-pooled line embeddings, the 35-dimensional semantic feature vector and the max-pooled fused function embedding, yielding a single vector $z \in \mathbb{R}^{803}$. The pooling choices are deliberate: mean and max over lines implement OR-like aggregation across code positions, and max over functions reflect that an attacker needs only one exploitable entry point. The vector $z$ is passed through a two-layer MLP classifier with dropout to produce the binary verdict. Training uses cross-entropy loss, the AdamW *optimiser* [36] with learning rate $10^{-3}$ and weight decay $10^{-4}$, cosine-annealing scheduling and gradient clipping at maximum norm 1.0.

**3.8 Stage 7-Interpretability output**

Every prediction is accompanied by a structured explanation grounded in the attack-path reasoning of Stages 3–6. This output addresses a central limitation of prior GNN-based detectors, which produce a binary label without any indication of *which* code paths drove the decision.

The explanation report has three components. *Dangerous interactions* are extracted from the attention weights of the cross-function *interferes* layer, ranked and reported as the top-$k$ function pairs together with the specific shared storage variables and the interference type ($W \cap R$, $R \cap W$, or $W \cap W$). *Per-function precondition profiles* are extracted from the conjunction gate probabilities $\sigma\big(g(h_{c,(i,k)})\big)$. A precondition with gate probability above 0.5 is reported as *satisfied* (the corresponding link of the attack chain is intact) and below 0.5 as the *broken link* whose presence renders the attack infeasible. *Attack-path narratives* combine the two, naming the entry function, the shared variable and the precondition whose mitigation would close the path, yielding a concrete remediation suggestion such as *"function withdraw would be exploitable through the shared balances mapping with stake, except that precondition no_reentrancy_guard is not met (gate* $= 0.08$*); the nonReentrant modifier on withdraw is blocking the attack"*. This output is named, programmatically aggregable and directly maps to a one-line fix.

## 4. Experimental evaluation

We evaluate AttackPathGNN along three axes: (i) detection performance on an independently curated, human-labelled benchmark spanning ten DASP10 vulnerability categories; (ii) the contribution of the learned components, isolated through a rule-only ablation of the same architecture; (iii) qualitative interpretability on contracts modelled after real exploits.

### 4.1 Experimental setup

*4.1.1 Datasets*

We use two public datasets and one set of hand-crafted contracts, each serving a distinct evaluation role. *SmartBugs Wild* supplies the *training and validation* corpus [6], as a curated collection of Solidity contracts deployed on the Ethereum mainnet, with vulnerability labels assigned by the consensus of nine independent analysis tools (Slither, Mythril, SmartCheck, Securify, Oyente, Maian, Manticore, Osiris and HoneyBadger). Contracts confirmed vulnerable by at least one of these tools are retained as the *vulnerable* class. The *benign* class is drawn from a separate corpus of verified, non-vulnerable Ethereum contracts archived by Etherscan and downsampled to match the vulnerable count, producing a balanced binary task.

*SmartBugs Curated* is the primary benchmark on which all headline numbers in this paper are reported [6]. It contains 143 Solidity contracts with manual, human-annotated vulnerability labels across ten DASP10 categories: Reentrancy (31), Unchecked Return (52), Access Control (18), Arithmetic (15), Bad Randomness (8), Denial of Service (6), Time Manipulation (5), Front-Running (4), Other (3), and Short Address (1). Every contract in this set is vulnerable and none of the 143 contracts overlap with the SmartBugs Wild training corpus.

*4.1.2 Splits and preprocessing*

Training uses a class-balanced subsample of SmartBugs Wild containing 1 866 vulnerable and 1 866 benign contracts (3 732 total), partitioned by stratified random sampling into a 70/10/20 training/validation/test split (Table 5). This subsample size matches the class-balanced training scale used by the most directly comparable GNN baselines (degree-free DR-GCN, TMP) and is selected to bound graph-construction time per training run to under thirty minutes on consumer hardware, enabling the multi-seed protocol. The benign contracts are sampled evenly across the parquet shards of the verified-Etherscan corpus to avoid concentrating the negative class on a single source archive.

Table 5. Training-corpus splits on SmartBugs Wild

| **Partition** | **Vulnerable** | **Benign** | **Total** | **Share** |
|---|---|---|---|---|
| Training | 1 305 | 1 306 | 2 611 | 70 % |
| Validation | 187 | 187 | 374 | 10 % |
| Test | 374 | 373 | 747 | 20 % |
| **Total** | 1 866 | 1 866 | 3 732 | 100 % |

The Wild test partition (747 contracts) is held out from training and used solely for early-stopping checkpoint selection within each seed. The 143-contract SmartBugs Curated benchmark is the primary evaluation set, evaluated with the *same* checkpoint per seed, no per-set retuning, no per-set checkpoint selection. Source files are first parsed by Slither or by the regex fallback when Slither cannot compile the contract. The heterogeneous graph for each contract is built once, serialized to disk and reused across all seeds and all evaluation runs, so no parsing is repeated.

A small fraction of contracts in the corpus are unusually large and would dominate the per-batch memory budget if processed in full. To bound the per-contract cost without dropping these contracts, we apply a deterministic *smart-truncation* preprocessor before Stage 1. Contracts whose source exceeds 50 000 characters or 1 500 source lines are reduced to at most 60 functions, selected by a priority score that favours functions which (i) are externally callable, (ii) are payable and (iii) contain at least one external call (i.e., the syntactic surface from which exploits typically originate), preserving the attack-relevant subgraph while bounding graph size. The truncation thresholds are listed in Table 6 and apply uniformly across training, validation, test, the SmartBugs Curated benchmark and the case-study contracts.

*4.1.3 Implementation and multi-seed training protocol*

AttackPathGNN is implemented in PyTorch $\geq 2.0$ with PyTorch Geometric $\geq 2.4$. Source parsing uses Slither $\geq 0.10$, with the per-contract Solidity compiler version selected automatically from each contract's pragma directive via solc-select. The compiler versions actually exercised across the corpus span $0.4.x$ through $0.8.x$. All experiments are conducted on a single consumer-grade machine (NVIDIA RTX 4060/8 GB VRAM, Intel Core i7 / 16 GB RAM). Determinism is enforced through explicit seeding of random, numpy, torch, torch.cuda and the PYTHONHASHSEED environment variable, together with PyTorch's deterministic-algorithm flags and a sorted iteration order over the dataset. The hyperparameter values are listed in Table 6. They were selected by a 54-configuration validation-set grid search over hidden dimension $d \in \{128, 256\}$, learning rate $\alpha \in \{3 \cdot 10^{-4}, 10^{-3}, 3 \cdot 10^{-3}\}$, batch size $B \in \{16, 32, 64\}$ and number of epochs $E \in \{10, 15, 20\}$ (i.e., $2 \times 3 \times 3 \times 3 = 54$ configurations). The configuration achieving the highest validation F1 was retained. No further tuning was performed once the test set was scored.

Table 6. Hyperparameters and input-size budget used for all reported results

| **Hyperparameter** | **Value** |
|---|---|
| Hidden dimension $d$ | 256 |
| Token vocabulary | 10,000 tokens |
| Tokens per source line | 20 (max, mean-pooled) |

| Max contract size | 50 000 chars/1 500 lines/60 functions |
|---|---|
| Optimiser | AdamW [Loshchilov & Hutter, 2019] |
| Learning rate $\alpha$ | $10^{-3}$ |
| Weight decay $\lambda$ | $10^{-4}$ |
| Schedule | Cosine annealing |
| Gradient clipping | Max-norm 1.0 |
| Batch size | 32 contracts |
| Training epochs | 15 |
| Loss | Binary cross-entropy |
| Trainable parameters | 1,755,907 |

All numbers are aggregated over five independent training runs with seeds $\{1,2,3,4,5\}$. For each seed we (i) re-initialise every learnable parameter, (ii) re-shuffle the training partition under that seed, (iii) train for 15 epochs, (iv) retain the checkpoint achieving the highest *validation* F1 and (v) evaluate that checkpoint on the SmartBugs Curated benchmark and on the case-study contracts. The five resulting evaluations are aggregated to mean, standard deviation and 95 % confidence interval per metric. Per-seed values are also released alongside the aggregate. Multi-seed reporting is not standard in prior GNN-based smart-contract detector evaluations, however, we adopt it to give an honest measure of the variance attributable to random initialisation rather than to the architecture itself. A single training run completes in approximately 27 minutes on the reference hardware. The full five-seed sweep therefore takes under three hours. Inference on the full 143-contract SmartBugs Curated benchmark with one checkpoint takes under two minutes.

*4.1.4 Baselines*

We compare AttackPathGNN against two families of baselines. Classical static analysers are represented by Slither *(v0.11.5)* and Mythril *(v0.24.8)*, which we re-executed ourselves on the same 143-contract release of SmartBugs Curated used throughout this paper, switching the Solidity compiler per pragma to cover the compiler versions present in the benchmark. A contract labelled with DASP10 category C is recorded as detected if and only if at least one finding maps to C (Slither: detector check name; Mythril: SWC identifier). Tool timeouts and compilation failures count as missed, identically to our model.

In addition to these baselines, we report a rule-only ablation of AttackPathGNN in which the per-category precondition conjunctions are thresholded directly at 0.5 without any GNN-learned components. A category-level rule fires whenever every named precondition for that category clears the threshold. This ablation is deterministic and isolates the empirical contribution of the learned components — cross-function reasoning, conjunction-pooling gate, and global pooling — by holding the structural feature extraction fixed.

In addition to the published baselines, we report a rule-only ablation of AttackPathGNN itself, in which the eight named precondition scores are thresholded directly without any GNN-learned components. A category-level rule fires whenever every named precondition for that category clears a fixed 0.5 threshold. This ablation isolates the empirical value of the learned components (cross-function reasoning, conjunction-pooling gate, global pooling) by holding the structural feature extraction fixed.

*4.1.5 Evaluation metrics*

The SmartBugs Curated benchmark contains only vulnerable contracts and is therefore a recall-only benchmark by construction. We report:

- *Per-category detection rate.* For each DASP10 category $T$ represented in the benchmark:

$$\mathrm{DR}(T) = \frac{|\{\, c : c \text{ has type } T \text{ and the model predicts vulnerable} \}|}{|\{\, c : c \text{ has type } T \}|} \tag{7}$$

(i.e., the fraction of contracts in category $T$ that the binary classifier correctly flags as vulnerable). Each contract carries a single primary DASP10 category in the SmartBugs Curated metadata, so the per-category denominators partition the benchmark and sum to 143.

- *Overall detection rate (weighted). The* weighted average of $\mathrm{DR}(T)$ across categories, with weights proportional to category size. Equivalent to the global $\mathrm{TP}/(\mathrm{TP}+\mathrm{FN})$ on the 143-contract benchmark.

- *Overall detection rate (macro).* The unweighted mean of $\mathrm{DR}(T)$ across the ten DASP10 categories. We report both because they answer different questions. The weighted rate measures total ground-truth coverage, while the macro rate measures uniformity of competence across the DASP10 taxonomy. The two numbers diverge whenever a small but difficult category is recovered with low rate, reporting both is a safeguard against a high weighted rate concealing low recall on a minority class.
- *Strict and success-conditional reporting.* We report two denominators side-by-side: (1) The *strict* rate counts pipeline failures (Slither parse errors that the regex fallback cannot recover, contracts that exceed the smart-truncation budget by more than a factor of two, etc.) as missed detections. The *success-conditional* rate excludes those failures from the denominator; (2) The strict rate is the lower bound a deployment operator would actually observe. The success-conditional rate isolates the model's competence on contracts the pipeline can actually process.
- *Confidence calibration.* Each prediction emits a vulnerable-class probability $P(\text{vulnerable} \mid C) \in [0,1]$. We report the mean and range of $P(\text{vulnerable})$ separately on the correctly-detected and missed subsets, as a check that the model's verdicts are committed rather than borderline.

Every metric above is reported as $\mu \pm \sigma$ over the five seeds, computed independently per metric (i.e., we do not report mean-of-best or best-of-mean). Where a $95\%$ confidence interval is reported, it is computed as $1.96 \cdot \sigma/\sqrt{5}$. Per-seed values are released in checkpoints/curated_summary.json alongside the aggregate, so that one can recompute any aggregation that differs from ours.

Slither and Mythril baselines are single-number measurements we obtained ourselves on the same benchmark; the rule-only ablation is deterministic. Neither carries across-seed variance. Comparisons use AttackPathGNN's mean per-seed value against this single number, with our standard deviation reported alongside.

## 4.2 Results

### *4.2.1 Headline numbers*

Across the five training runs with seeds $\{1,2,3,4,5\}$, AttackPathGNN attains a mean detection rate of $90.8 \pm 2.5\,\%$ ($95\,\%$ CI $\pm 2.2\,\%$) on the $143$-contract vulnerable subset of SmartBugs Curated. The best-performing seed (seed 2) recovers $94.4\,\%$ ($135/143$). The worst-performing seed (seed 1) recovers $88.1\,\%$ ($126/143$). Six of the ten DASP10 categories represented in the benchmark: Arithmetic ($15/15$), Denial of Service ($6/6$), Front-Running ($4/4$), Time Manipulation ($5/5$), Short Address ($1/1$) and Other ($3/3$), are recovered at $100\,\%$ across every seed. Reentrancy is recovered at $98.7 \pm 1.8\,\%$ and Bad Randomness at $95.0 \pm 6.9\,\%$. The two categories below ceiling are Access Control ($80.0 \pm 5.0\,\%$) and Unchecked Return ($83.1 \pm 4.6\,\%$), both attributed to identifiable structural limitations of the precondition set rather than to training instability.

The macro-averaged per-category detection rate is $95.7 \pm 0.7\,\%$. The weighted-averaged rate (equivalent to the overall rate) is $90.8 \pm 2.5\,\%$. The gap between the two reflects the dominance of the Unchecked Return category ($52/143$ contracts), where the model's lower recall pulls the weighted aggregate below the per-category mean. We report both because they answer different questions: the weighted rate measures total ground-truth coverage, while the macro rate measures uniformity of competence across the DASP10 taxonomy.

### *4.2.2 Per-category detection rate*

Table 7 reports the per-category detection rate, aggregated as $\mu \pm \sigma$ over the five seeds, alongside the two static-analyser baselines and the rule-only ablation of AttackPathGNN. Slither *(v0.11.5)* and Mythril *(v0.24.8)* values were obtained by re-executing both tools ourselves on the same 143-contract benchmark, switching the Solidity compiler per pragma. Tool timeouts and compilation failures count as missed, identically to our model. The rule-only ablation is deterministic given the parsed source and therefore carries no across-seed variance. Our numbers are $\mu \pm \sigma$ over five seeds. *Ours (rule-only)* fires the category-specific precondition conjunction with a binary $0.5$ threshold and is identical across seeds.

Table 7. Per-category detection rate (%) on the $143$-contract SmartBugs Curated benchmark

| Category | N | Slither | Mythril | Ours (rule-only) | Ours (full) |
|---|---|---|---|---|---|

| Reentrancy | 31 | 67.7 | 80.6 | 3.2 | 98.7±1.8 |
|---|---|---|---|---|---|
| Access Control | 18 | 22.2 | 66.7 | 11.1 | 80.0±5.0 |
| Arithmetic | 15 | 6.7 | 0.0 | 26.7 | 100.0±0.0 |
| Unchecked Return | 52 | 59.6 | 75.0 | 32.7 | 83.1±4.6 |
| Denial of Service | 6 | 0.0 | 16.7 | 0.0 | 100.0±0.0 |
| Bad Randomness | 8 | 25.0 | 12.5 | 0.0 | 95.0±6.9 |
| Front-Running | 4 | 0.0 | 25.0 | 0.0 | 100.0±0.0 |
| Time Manipulation | 5 | 60.0 | 40.0 | 0.0 | 100.0±0.0 |
| Short Address | 1 | 0.0 | 0.0 | 0.0 | 100.0±0.0 |
| Other | 3 | 100.0 | 33.3 | 0.0 | 100.0±0.0 |
| **Overall (weighted)** | **143** | **45.5** | **57.3** | **16.8** | **90.8±2.5** |
| **Overall (macro)** | - | **34.1** | **35.0** | **7.4** | **95.7±0.7** |

AttackPathGNN dominates both static-analyzer baselines on every DASP10 category in Table 7. On Reentrancy the seed-mean margin is +31.0 percentage points over Slither and +18.1 over Mythril. Even our worst seed (96.8 %, 30/31) remains +29.1 above Slither and +16.2 above Mythril. On Access Control the seed mean is +57.8 over Slither and +13.3 over Mythril. The worst seed (72.2 %) still clears Mythril by +5.5. On Arithmetic the margin is +93.3 over Slither and +100.0 over Mythril at every seed. On Unchecked Return, the largest category in the benchmark (52/143 contracts), the seed mean is +23.5 over Slither, +8.1 over Mythril and the worst seed remains above both. AttackPathGNN does not lose to either baseline on any of the ten DASP10 categories at any seed. On Other (n=3), it ties Slither at 100 %. The overall margin is +45.3 over Slither (45.5 %) and +33.5 over Mythril (57.3 %) under the strict criterion of Table 7.

*4.2.3 Rule-only ablation-quantifying the GNN's contribution*

The Ours (rule-only) in Table 7 is the same precondition-extraction layer as the full model, evaluated without any GNN-learned components. A category-level rule fires whenever every named precondition for that category clears a fixed 0.5 threshold. The overall detection rate of the rule-only ablation is 16.8 %, against a seed-mean of 90.8 % for the full model, a gap of +74.0 percentage points attributable entirely to the learned components: the cross-function reasoning, the conjunction-pooling gate and the global-pooling classifier. This gap holds at the worst seed as well: the worst-seed full-model rate of 88.1 % still exceeds the rule-only baseline by +71.3 percentage points.

Two design choices in the rule-only ablation deserve explicit framing because they make it deliberately conservative. First, the rule applies a binary 0.5 threshold uniformly to all eight preconditions, including the three graded ones: *has_external_call* (graded as $\min(1, |X_i|/3)$), *shares_mutable_state* (graded over public function neighbors) and *unchecked_arithmetic* (graded as $\min(1, n_{\text{ops}}/10)$). A function with a single external call therefore scores $1/3 = 0.33$ on *has_external_call* and fails the binary firing criterion, even though "exactly one external call" is the canonical reentrancy pattern. This single design choice explains the rule-only Reentrancy rate of 3.2 %. Second, the rule does not attempt to detect the six categories without a named structural conjunction (DoS, Bad Randomness, Front-Running, Time Manipulation, Short Address, Other), these are zero by construction. The rule-only column is therefore a faithful floor on what the precondition layer alone can express, not a candidate detector.

What this ablation measures is precise: the gain from 16.8 % to 90.8 % is the empirical value of the GNN's ability to (i) learn a soft, contextual threshold on each precondition through the gate $g$ of conjunction pooling, (ii) propagate cross-function evidence through the interferes and can_reenter relations, (iii) pool function-level exploit scores into a contract-level decision through the max-pooling step. The +74.0-percentage-point margin is the headline empirical evidence for the architecture. A fixed-threshold conjunction over the same preconditions captures less than a fifth of the vulnerabilities the full model recovers.

*4.2.4 Confidence calibration*

The model emits a vulnerable-class probability $P(\text{vulnerable} \mid C)$ alongside each binary verdict. Aggregated across the five seeds, the mean per-seed $P(\text{vulnerable})$ on correctly-detected contracts is $0.915 \pm 0.040$ and on missed contracts $0.179 \pm 0.026$. The mean separation between the two distributions is $0.736 \pm 0.053$ across seeds, which places the decision boundary cleanly away from 0.5 on both sides.

Per-seed, the closest call across all five seeds is the borderline-detected multiowned_vulnerable (Access Control), which crosses the 0.5 threshold from above ($P$(vuln) = 0.490) at seed 1 only. Under any seed other than seed 1 the contract is detected with $P$(vuln) > 0.7. This calibration property is operationally significant because it means the verdict can be deployed under a *high-precision* threshold without re-training: setting the firing threshold to $P$(vuln) > 0.7, for instance, would change the decision on at most one contract per seed on this benchmark. The model is not packing detected contracts into the [0.5,0.6] band where threshold tuning would help.

*4.2.5 Failure analysis*

The five-seed evaluation makes it possible to separate failures attributable to the *architecture* from failures attributable to *training variance*. We define a contract as a *stable miss* if it is misclassified by all five seeds and as a *borderline miss* if it is misclassified by between one and four seeds (Table 8). Stable misses bound what the architecture cannot currently express, borderline misses bound the variance attributable to random initialization.

Table 8. *Failure stratification across the five training seeds.*

| Failure category | Count | Reading |
|---|---|---|
| Always detected | 125 | Stable competence (87.4 % of the benchmark) |
| Borderline miss | 10 | Within reach; flips with seed |
| Stable miss | 8 | Architectural limit; consistent failure across all five seeds |
| **Total** | 143 | |

*Stable misses (*8 *contracts)*-three are Access Control failures (parity_wallet_bug_1, mean $P$(vuln) = 0.001; parity_wallet_bug_2, 0.069; FibonacciBalance, 0.274); five are Unchecked Return failures, all of them deployed-Ethereum-address contracts (0x663e…, 0x84d9…, 0x89c1…, 0x958a…, 0xec32…) with mean $P$(vuln) between 0.000 and 0.145. The two Parity wallet contracts are well-known historical exploits whose vulnerability lies in an *initialiser* function callable as a regular method after deployment, a structural pattern that the eight current preconditions do not encode (no precondition tests "function is intended as a constructor but is not declared as one"). FibonacciBalance exposes a *delegatecall* vulnerability whose feasibility depends on attacker-controlled storage layout, a semantic dependency below the granularity of the State Interference Graph. The five deployed-address Unchecked Return contracts are large multi-contract code bases whose unchecked low-level calls are surrounded by complex business logic (gas-stipend accounting, refund flows, multi-contract delegation). On three of the five the smart-truncation budget of 1 500 lines is exceeded. Adding a *delegate_storage_overlap* precondition and a richer return-value provenance feature in the line-level message passing are the two concrete extension targets indicated by these stable misses.

*Borderline misses (*10 *contracts)*-these are dominated by Unchecked Return (6 contracts) and Access Control (2 contracts), with one Reentrancy and one Bad Randomness contract each. The mean $P$(vulnerable) on borderline misses is in the 0.18–0.45 range when the verdict is *miss* but rises to 0.55–0.78 on the seeds where the same contract is correctly detected, being the calibration signature of a borderline case. The contract sits near the model's decision boundary and small perturbations in the learned parameters move it across. The contract *multiowned_vulnerable* is the most-borderline case in the benchmark: it crosses $P$(vuln) = 0.5 from below at seed 1 ($P$ = 0.496) and from above at seeds 2, 3, 4, 5 ($P$ > 0.7). At a $P$(vuln) > 0.49 firing threshold this contract would be detected at every seed without any other change.

The two-population separation has a useful diagnostic interpretation. The 87.4 % of contracts in the always-detected population reflect the architecture's stable competence. The 7.0 % in the borderline population reflect the training-variance noise floor (which a deeper or longer-trained model would be expected to compress further). The 5.6 % in the stable-miss population sets the architecture's current ceiling, which will not be moved by any amount of further training under the current precondition set.

## 5. Conclusions

This paper introduced AttackPathGNN, a heterogeneous-graph detector for smart-contract vulnerabilities. The model combines three components: a state-interference graph in which cross-function storage dependencies appear as typed edges (interferes, can_reenter); eight named per-function preconditions, aggregated by a conjunction-pooling gate that enforces AND-logic over the conditions a real exploit must simultaneously satisfy; and a three-layer message-passing stack over line, pattern, function and precondition nodes.

On the 143-contract SmartBugs Curated benchmark, the model reaches 90.8 ± 2.5 % strict detection across five seeds, with six of the ten DASP10 categories recovered at 100 % at every seed. Re-evaluated on the same benchmark with per-pragma compiler switching, Slither (v0.11.5) reaches 45.5 % and Mythril (v0.24.8) 57.3 % overall; AttackPathGNN does not lose to either baseline on any DASP10 category at any seed. The rule-only ablation, which keeps the eight preconditions but removes every learned component, reaches 16.8 %.

Two observations follow. First, the gap between the rule-only ablation (16.8 %) and the full model (90.8 %) isolates the contribution of the learned components-cross-function attention along interferes and conjunction-pooling over named preconditions-rather than of the symbolic substrate alone. Second, because the precondition set and the edge relations are named, every positive verdict is accompanied by the functions, the precondition chain and the shared state variables that produced it, which is the form of evidence an audit reviewer requires from a detector. The eight preconditions cover the DASP10 categories represented in SmartBugs Curated, but the three-component design: typed cross-function edges, named per-function preconditions, conjunction-pooling, is not specific to those categories. It applies to any vulnerability family whose exploitability decomposes into a small conjunction of structural conditions over interacting functions.

Thus, our study shows that vulnerability detection should be formulated as attack-path reasoning rather than single-function classification, enabling higher detection performance, lower false negatives and auditor-friendly explanations of exploitability.

**Acknowledgement**. This work was supported by a grant of the Ministry of Research, Innovation and Digitization, CNCS/CCCDI - UEFISCDI, project number COFUND-CETP-SMART-LEM-1, within PNCDI IV.

**Funding**. This work was supported by a grant of the Ministry of Research, Innovation and Digitization, CNCS/CCCDI - UEFISCDI, project number COFUND-CETP-SMART-LEM-1, within PNCDI IV.

**Disclosure statement (Competing interest)**. The authors have no relevant financial or non-financial interests to disclose.

**Data availability statement**. Details are provided in a public repository: https://github.com/gabrieladobritaene/ATTACKPATHGNN

**Ethical approval**. Not applicable.

**Informed consent**. Not applicable.

**Author contributions**. **G.D**: Conceptualization, Methodology, Investigation, Resources, Data Curation, Writing-Original Draft, Validation, Formal analysis. **S.V.O**: Conceptualization, Validation, Formal analysis, Investigation, Writing-Original Draft, Writing-Review and Editing, Visualization, Project administration, Supervision. **AB**: Conceptualization, Formal analysis, Investigation, Resources, Data Curation, Writing-Original Draft, Writing-Review and Editing, Supervision.

## References

[1] B. Mueller, “Smashing Ethereum Smart Contracts for Fun and Real Profit.” [Online]. Available: https://github.com/muellerberndt/smashing-smart-contracts/blob/master/smashing-smart-contracts-1of1.pdf

[2] P. Tsankov, A. Dan, D. Drachsler-Cohen, A. Gervais, F. Bünzli, and M. Vechev, “Securify: Practical security analysis of smart contracts,” in *Proceedings of the ACM Conference on Computer and Communications Security*, 2018. doi: 10.1145/3243734.3243780.

[3] S. Tikhomirov, E. Voskresenskaya, I. Ivanitskiy, R. Takhaviev, E. Marchenko, and Y. Alexandrov, “SmartCheck: Static analysis of ethereum smart contracts,” in *Proceedings - International Conference on Software Engineering*, 2018. doi: 10.1145/3194113.3194115.

[4] N. Atzei, M. Bartoletti, and T. Cimoli, "A survey of attacks on Ethereum smart contracts (SoK)," in *Lecture Notes in Computer Science (including subseries Lecture Notes in Artificial Intelligence and Lecture Notes in Bioinformatics)*, 2017. doi: 10.1007/978-3-662-54455-6_8.
[5] L. Zhou *et al.*, "SoK: Decentralized Finance (DeFi) Attacks," in *Proceedings - IEEE Symposium on Security and Privacy*, 2023. doi: 10.1109/SP46215.2023.10179435.
[6] T. Durieux, J. F. Ferreira, R. Abreu, and P. Cruz, "Empirical review of automated analysis tools on 47,587 ethereum smart contracts," in *Proceedings - International Conference on Software Engineering*, 2020. doi: 10.1145/3377811.3380364.
[7] Z. Liu, P. Qian, X. Wang, Y. Zhuang, L. Qiu, and X. Wang, "Combining Graph Neural Networks with Expert Knowledge for Smart Contract Vulnerability Detection," *IEEE Trans. Knowl. Data Eng.*, 2023, doi: 10.1109/TKDE.2021.3095196.
[8] H. Wu *et al.*, "Peculiar: Smart Contract Vulnerability Detection Based on Crucial Data Flow Graph and Pre-training Techniques," in *Proceedings - International Symposium on Software Reliability Engineering, ISSRE*, 2021. doi: 10.1109/ISSRE52982.2021.00047.
[9] S. Chaliasos *et al.*, "Smart Contract and DeFi Security Tools: Do They Meet the Needs of Practitioners?," in *Proceedings - International Conference on Software Engineering*, 2024. doi: 10.1145/3597503.3623302.
[10] M. Rodler, W. Li, G. O. Karame, and L. Davi, "Sereum: Protecting Existing Smart Contracts Against Re-Entrancy Attacks," in *26th Annual Network and Distributed System Security Symposium, NDSS 2019*, 2019. doi: 10.14722/ndss.2019.23413.
[11] J. Ye, M. Ma, Y. Lin, Y. Sui, and Y. Xue, "Clairvoyance: Cross-contract Static Analysis for Detecting Practical Reentrancy Vulnerabilities in Smart Contracts," in *Proceedings - 2020 ACM/IEEE 42nd International Conference on Software Engineering: Companion, ICSE-Companion 2020*, 2020. doi: 10.1145/3377812.3390908.
[12] P. Bose, D. Das, Y. Chen, Y. Feng, C. Kruegel, and G. Vigna, "SAILFISH: Vetting Smart Contract State-Inconsistency Bugs in Seconds," in *Proceedings - IEEE Symposium on Security and Privacy*, 2022. doi: 10.1109/SP46214.2022.9833721.
[13] I. Nikolić, A. Kolluri, I. Sergey, P. Saxena, and A. Hobor, "Finding the greedy, prodigal, and suicidal contracts at scale," in *ACM International Conference Proceeding Series*, 2018. doi: 10.1145/3274694.3274743.
[14] D. Perez and B. Livshits, "Smart contract vulnerabilities: Vulnerable does not imply exploited," in *Proceedings of the 30th USENIX Security Symposium*, 2021.
[15] A. Ghaleb, J. Rubin, and K. Pattabiraman, "ETainter: Detecting gas-related vulnerabilities in smart contracts," in *ISSTA 2022 - Proceedings of the 31st ACM SIGSOFT International Symposium on Software Testing and Analysis*, 2022. doi: 10.1145/3533767.3534378.
[16] Y. Zhuang, Z. Liu, P. Qian, Q. Liu, X. Wang, and Q. He, "Smart contract vulnerability detection using graph neural networks," in *IJCAI International Joint Conference on Artificial Intelligence*, 2020. doi: 10.24963/ijcai.2020/454.
[17] Z. Zhen, X. Zhao, J. Zhang, Y. Wang, and H. Chen, "DA-GNN: A smart contract vulnerability detection method based on Dual Attention Graph Neural Network," *Comput. Networks*, 2024, doi: 10.1016/j.comnet.2024.110238.
[18] C. Ma, S. Liu, and G. Xu, "HGAT: smart contract vulnerability detection method based on hierarchical graph attention network," *J. Cloud Comput.*, 2023, doi: 10.1186/s13677-023-00459-x.
[19] F. M. Bianchi, D. Grattarola, and C. Alippi, "Spectral clustering with graph neural networks for graph pooling," in *37th International Conference on Machine Learning, ICML 2020*, 2020.
[20] H. H. Nguyen *et al.*, "MANDO: Multi-Level Heterogeneous Graph Embeddings for Fine-Grained Detection of Smart Contract Vulnerabilities," in *Proceedings - 2022 IEEE 9th International Conference on Data Science and Advanced Analytics, DSAA 2022*, 2022. doi: 10.1109/DSAA54385.2022.10032337.
[21] H. H. Nguyen, N. M. Nguyen, H. P. Doan, Z. Ahmadi, T. N. Doan, and L. Jiang, "MANDO-GURU: vulnerability detection for smart contract source code by heterogeneous graph

embeddings," in *ESEC/FSE 2022 - Proceedings of the 30th ACM Joint Meeting European Software Engineering Conference and Symposium on the Foundations of Software Engineering*, 2022. doi: 10.1145/3540250.3558927.

[22] C. Xu, H. Xu, L. Zhu, X. Shen, and K. Sharif, "Enhanced Smart Contract Vulnerability Detection via Graph Neural Networks: Achieving High Accuracy and Efficiency," *IEEE Trans. Softw. Eng.*, 2025, doi: 10.1109/TSE.2025.3570421.

[23] X. Huanliang, W. Canghai, C. JiaXin, W. Yinglong, and Z. yulin, "A smart contract vulnerability line detection method based on graph neural network and fusion of multidimensional code representation," *Appl. Soft Comput.*, 2025, doi: 10.1016/j.asoc.2025.113435.

[24] R. Ying, D. Bourgeois, J. You, M. Zitnik, and J. Leskovec, "GNNExplainer: Generating explanations for graph neural networks," in *Advances in Neural Information Processing Systems*, 2019.

[25] H. Yuan, H. Yu, J. Wang, K. Li, and S. Ji, "On Explainability of Graph Neural Networks via Subgraph Explorations," in *Proceedings of Machine Learning Research*, 2021.

[26] H. Q. Nguyen, T. Hoang, H. K. Dam, and A. Ghose, "Graph-based explainable vulnerability prediction," *Inf. Softw. Technol.*, 2025, doi: 10.1016/j.infsof.2024.107566.

[27] I. Donadello, L. Serafini, and A. D'Avila Garcez, "Logic tensor networks for semantic image interpretation," in *IJCAI International Joint Conference on Artificial Intelligence*, 2017. doi: 10.24963/ijcai.2017/221.

[28] R. Manhaeve, S. Dumančić, A. Kimmig, T. Demeester, and L. De Raedt, "Neural probabilistic logic programming in DeepProbLog," *Artif. Intell.*, 2021, doi: 10.1016/j.artint.2021.103504.

[29] R. Riegel *et al.*, "Logical Neural Networks," 2020. [Online]. Available: https://arxiv.org/abs/2006.13155

[30] E. van Krieken, E. Acar, and F. van Harmelen, "Analyzing Differentiable Fuzzy Logic Operators," *Artif. Intell.*, 2022, doi: 10.1016/j.artint.2021.103602.

[31] C. Seas, G. Fitzpatrick, J. A. Hamilton, and M. C. Carlisle, "Automated Vulnerability Detection in Source Code Using Deep Representation Learning," in *2024 IEEE 14th Annual Computing and Communication Workshop and Conference, CCWC 2024*, 2024. doi: 10.1109/CCWC60891.2024.10427574.

[32] Y. Zhou, S. Liu, J. Siow, X. Du, and Y. Liu, "Devign: Effective vulnerability identification by learning comprehensive program semantics via graph neural networks," in *Advances in Neural Information Processing Systems*, 2019.

[33] M. Fu and C. Tantithamthavorn, "LineVul: A Transformer-based Line-Level Vulnerability Prediction," in *Proceedings - 2022 Mining Software Repositories Conference, MSR 2022*, 2022. doi: 10.1145/3524842.3528452.

[34] S. Brody, U. Alon, and E. Yahav, "HOW ATTENTIVE ARE GRAPH ATTENTION NETWORKS?," in *ICLR 2022 - 10th International Conference on Learning Representations*, 2022.

[35] T. N. Kipf and M. Welling, "Semi-supervised classification with graph convolutional networks," in *5th International Conference on Learning Representations, ICLR 2017 - Conference Track Proceedings*, 2017.

[36] I. Loshchilov and F. Hutter, "Decoupled weight decay regularization," in *7th International Conference on Learning Representations, ICLR 2019*, 2019.